\documentclass[final]{svjour3}
\usepackage{graphicx}
\usepackage{rotating}
\usepackage{amssymb}
\usepackage{mathptmx}
\usepackage[numbers]{natbib}
\makeatletter
\journalname{Journal of Low Temperature Physics}

\bibpunct{}{}{,}{s}{}{,}

\begin{document}

\newcommand{\hdblarrow}{H\makebox[0.9ex][l]{$\downdownarrows$}-}
\title{Transmission and Reflection of Collective Modes in Spin-1 Bose-Einstein Condensate}

\author{Shohei Watabe$^1$ \and Yusuke Kato$^2$}

\institute{1: Institute of Physics, Department of Physics, The University of Tokyo, Komaba 3-8-1 Meguro-ku, Tokyo, 153-8902, Japan
\\
\email{watabe@vortex.c.u-tokyo.ac.jp}
\\
2: Department of Basic Science, The University of Tokyo, Komaba 3-8-1 Meguro-ku, Tokyo, 153-8902, Japan
\email{yusuke@phys.c.u-tokyo.ac.jp}}

\date{June.15.2009}

\maketitle

\keywords{quantum gas, spin-1 BEC, excitation, tunneling problem}

\begin{abstract}
We study tunneling properties of collective excitations in spin-1 Bose-Einstein condensates. 
In the absence of magnetic fields, 
the total transmission in the long wavelength limit 
occurs in all kinds of excitations but the quadrupolar spin mode in the ferromagnetic state. 
The quadrupolar spin mode alone shows the total reflection. 
A difference between those excitations comes from 
whether the wavefunction of an excitation corresponds to that of the condensate in the long wavelength limit. 
The correspondence results in the total transmission as in the spinless BEC. 

PACS numbers: 03.75.Lm, 03.75.Mn
\end{abstract}

\section{Introduction} 
Bose-Einstein condensates (BECs) with internal degrees of freedom are studied 
as realistic objects in the field of ultracold atomic gases~\cite{Stenger1998, Higbie2005, Sadler2006}. 
As contrasted to a spinless BEC, 
a spinor BEC has rich ground states, and also has many types of excitations. 
We thus expect to find new phenomena in the spinor BEC, not seen in the spinless BEC. 
Through studies of these spinor BECs, 
features of the spinless BEC will be also clear. 

One of the striking properties in the spinless BEC 
is the total transmission of the Bogoliubov excitation in the low-energy limit~\cite{Kovrizhin2001,Kagan2003}. 
This problem has been extensively and intensively studied, 
and many studies have revealed basic and unknown properties 
of the spinless BEC~\cite{Kovrizhin2001, Kagan2003, Danshita2006, Kato2007, Watabe2008, Ohashi2008, Takahashi2009}. 
However, tunneling properties of excitations in the spin-1 BEC have been untouched in earlier papers. 

In the spin-1 BEC, there are 2-types of the ground state, called ferromagnetic state and polar state~\cite{Ohmi1998, Ho1998}. 
For excitations, there are 3-types in each ground state, 
such as a Bogoliubov mode and spin modes. 
These excitations have different wavefunctions and dispersion relations. 
A natural question concerning the spin-1 BEC is what kinds of excitations transmit perfectly, transmit partially or reflect perfectly. 
The aim of this study is 
to reveal transmission properties of those excitations in the spin-1 BEC. 
The results lead to a further understanding of magnetism and transmission of spin excitations 
in spinor BECs. 

Results are summarized as follows. 
We find that 
the total transmission occurs in all kinds of excitations but the quadrupolar spin mode in the ferromagnetic state. 
Wavefunctions of these excitations correspond to the condensate wavefunction in the long wavelength limit, 
as seen in the spinless BEC~\cite{Kato2007}. 
We also find that the quadrupolar spin mode does not show this correspondence, and experiences the total reflection.

\section{Formulation} 
We start with the static Gross-Pitaevskii (GP) equations for the spin-1 BEC system~\cite{Ohmi1998, Ho1998} 
for the order parameters $\Phi_{i}$ with spin index $i=\pm1, 0$. 
These equations in a simplified form~\cite{Saito2006} are given by 
\begin{eqnarray}
\left ( -\frac{\hbar^{2}}{2m} \nabla^{2} + V_{\rm ext} -\mu + c_{0} n \mp g \mu_{\rm B} B^{z} \right  ) \Phi_{\pm 1} 
+ c_{1}(\frac{1}{\sqrt{2}} F_{\mp} \Phi_{0} \pm F_{z}\Phi_{\pm1}) = 0, 
\\
\left (-\frac{\hbar^{2}}{2m} \nabla^{2} + V_{\rm ext} -\mu + c_{0} n \right )\Phi_{0} 
+ \frac{c_{1}}{\sqrt{2}}(F_{+} \Phi_{+1} + F_{-}\Phi_{-1}) = 0, 
\end{eqnarray}
where $n = \sum\limits_{i}|\Phi_{i}|^{2}$, $F_{z} = |\Phi_{+1}|^{2} - |\Phi_{-1}|^{2}$, 
and $F_{+} = F_{-}^{*} = \sqrt{2} (\Phi_{+1}^{*}\Phi_{0} + \Phi_{0}^{*}\Phi_{-1})$. 
Interaction strengths are given by $c_{0} = 4\pi\hbar^{2}(2a_{2} + a_{0})/(3m)$, and 
$c_{1} = 4\pi\hbar^{2}(a_{2} - a_{0})/(3m)$, 
where $m$ is a mass, and $a_{S}$ is the $s$-wave scattering length with collision process with total spin $S$. 
$g$ is the Land\`e's $g$-factor, $\mu_{\rm B}$ is the Bohr magneton. 
In the case $c_{1} < 0$, the ground state is so called the ferromagnetic state~\cite{Ohmi1998, Ho1998}, 
while the ground state with $c_{1} > 0$ is the polar state~\cite{Ohmi1998, Ho1998}. 
In this article, we discuss the tunneling problem of excitations, 
and hence we shall consider small fluctuations $\delta\Phi_{i}$ 
from the condensate wavefunctions: $\Phi_{i} + \delta\Phi_{i}$. 

In a homogeneous system, 
energies and wavefunctions of excitations have been obtained~\cite{Ohmi1998, Ho1998}. 
We summarize the physical properties of excitations in the spin-1 BEC. 
In the ferromagnetic state, one mode is the Bogoliubov mode, 
where the fluctuation is $\delta\Phi_{+1}$ and the excitation spectrum is $E = \sqrt{\varepsilon [\varepsilon + 2(c_{0}+c_{1})n]}$. 
$\varepsilon$ is given by $\varepsilon = \hbar^{2}k^{2}/(2m)$. 
Another mode is a spin wave, where the fluctuation is $\delta\Phi_{0}$ 
and the excitation energy is given by $E = \varepsilon + g \mu_{\rm B}B^{z}$. 
The other mode is a quadrupolar spin mode~\cite{Ho1998}, 
where the fluctuation is $\delta\Phi_{-1}$ with the energy $E = \varepsilon + 2|c_{1}|n + 2 g \mu_{\rm B}B^{z}$. 

In the polar state in the absence of a magnetic field $(B^{z} = 0)$, 
one mode is the Bogoliubov mode, where the fluctuation is $\delta\Phi_{0}$ 
with $E = \sqrt{\varepsilon (\varepsilon + 2c_{0}n)}$. 
The others are spin wave modes doubly degenerated, where fluctuations are $\delta\Phi_{\pm1}$ with $E = \sqrt{\varepsilon (\varepsilon + 2c_{1}n)}$. 

We calculate the transmission coefficient of those excitations against a potential barrier in the following way. 
First, we solve the above Gross-Pitaevskii equations including a potential barrier, 
$V_{\rm ext}(x) = V_{0} \theta(a-|x|)$, with the Heaviside's step function $\theta (x)$. 
Using the resulting order parameters, we evaluate transmission and reflection coefficients by solving equations for small fluctuations,
such as a Bogoliubov equation, 
with imposing the boundary conditions where incident, reflected and transmitted waves exist at asymptotic regimes. 
The details of formulation will be reported in another article. 

\section{Results} 
\subsection{Ferromagnetic state}
We first show the results in the ferromagnetic state. 
We use parameters $a_{0}$ and $a_{2}$ as $a_{0} : a_{2} = 110: 107$ as in Ref. 2. 
Figures 1 and 2 show transmission coefficients and phase shifts of excitations, respectively. 
The phase shift is defined by an argument of an amplitude transmission coefficient. 
Figures (a), (b) and (c) show the results for the Bogoliubov mode, spin wave mode, and quadrupolar spin mode, 
respectively. 

Note that all the excitations but the fluctuation of the quadrupolar spin show the total transmission 
in the long wavelength limit. 
Phase shifts of these two excitations vanish in this limit. 
These behaviors are qualitatively the same as those in the spinless BEC, irrespective of dispersion relation. 
The excitation of the quadrupolar spin alone shows the total reflection, and this phase shift does not vanish in the long wavelength limit. 
We found that the argument of the amplitude reflection coefficient is $\pi$ in the long wavelength limit, 
corresponding to the fixed-end reflection. 
We have checked that the results in the presence of the magnetic field $(B^{z} \neq 0)$ 
are qualitatively the same as ones at $B^{z} = 0$. 
\begin{figure}
\begin{center}
\includegraphics[width=0.9\linewidth,keepaspectratio]{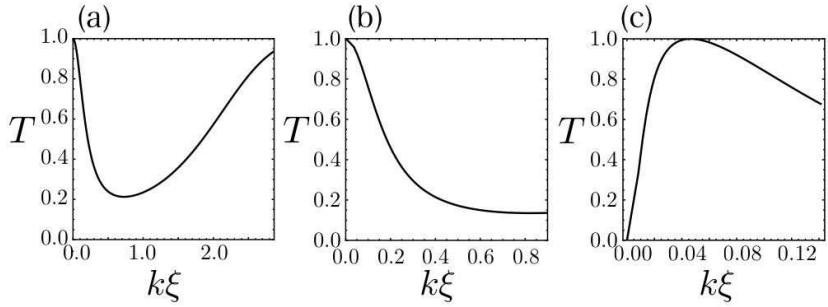}
\end{center}
\caption{Transmission coefficients of excitations in the ferromagnetic state as a function of the wavenumber at $B^{z}=0$. 
Each figure corresponds to (a) Bogoliubov excitation, (b) spin wave mode, and (c) quadrupolar spin mode. 
We set parameters $V_{0} = 3 c_{0}n$ and $a = \xi/2$, where $\xi \equiv \sqrt{2mc_{0}n}/\hbar$. 
}
\label{fing1}
\end{figure}
\begin{figure}
\begin{center}
\includegraphics[width=0.9\linewidth,keepaspectratio]{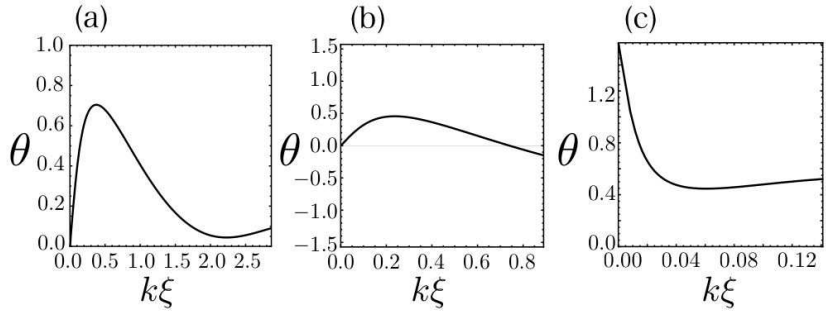}
\end{center}
\caption{Phase shifts of excitations in the ferromagnetic state as a function of the wavenumber at $B^{z}=0$. 
Each figure corresponds to (a) Bogoliubov excitation, (b) spin wave mode, and (c) quadrupolar spin mode. 
We set parameters $V_{0} = 3 c_{0}n$ and $a = \xi/2$. }
\label{fing2}
\end{figure}
\subsection{Polar state}
\begin{figure}
\begin{center}
\includegraphics[width=0.65\linewidth,keepaspectratio]{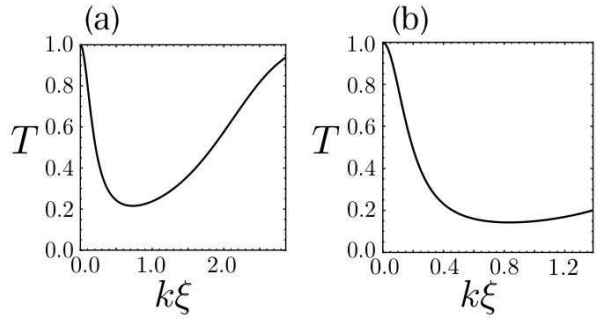}
\end{center}
\caption{Transmission coefficients of excitations in the polar state as a function of the wavenumber at $B^{z} = 0$. 
Each figure corresponds to (a) Bogoliubov excitation, and (b) spin wave modes doubly degenerated. 
We use parameters $V_{0} = 3 c_{0}n$ and $a = \xi/2$. }
\label{fig3}
\end{figure}
\begin{figure}
\begin{center}
\includegraphics[width=0.65\linewidth,keepaspectratio]{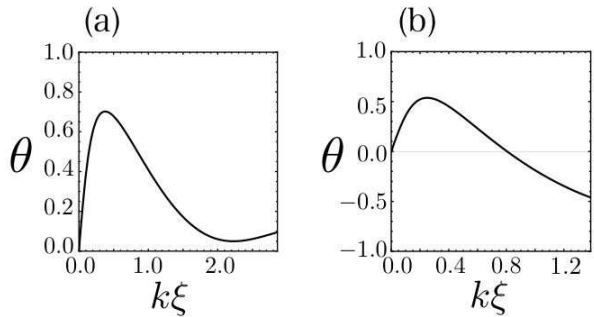}
\end{center}
\caption{Phase shifts of excitations in the polar state as a function of the wavenumber at $B^{z} = 0$. 
Each figure corresponds to (a) Bogoliubov excitation, and (b) spin wave modes doubly degenerated. 
We set parameters $V_{0} = 3 c_{0}n$ and $a = \xi/2$. }
\label{fig4}
\end{figure}
We present the results in the polar state at $B^{z} = 0$. 
We use parameters $a_{0}$ and $a_{2}$ as $a_{0} : a_{2} = 46 : 52$ as in Ref. 2. 
Figures 3 and 4 show transmission coefficients and phase shifts of excitations, respectively. 
In each figure, (a) and (b) show the results of the Bogoliubov excitation, and spin wave modes doubly degenerated, respectively. 
The total transmissions in the long wavelength limit occur in all excitations 
and the phase shifts vanish in the same limit. 

We note that, in the polar state at $B^{z} \neq 0$, 
there is a special case $c_{0} = c_{1}$, 
which corresponds to the integrable condition found by Ieda~\cite{2004Ieda}. 
In this condition, excitations of $\pm 1$ components from the ground state are completely decoupled, 
and hence the properties of excitations are the same as those in the spinless BEC. 
We confirmed 
the total transmission of 3-types of excitations in the long wavelength limit under the condition $c_{0} = c_{1}$. 

\section{Discussion}
\begin{figure}[htbp]
\begin{center}
\includegraphics[width=0.8\linewidth,keepaspectratio]{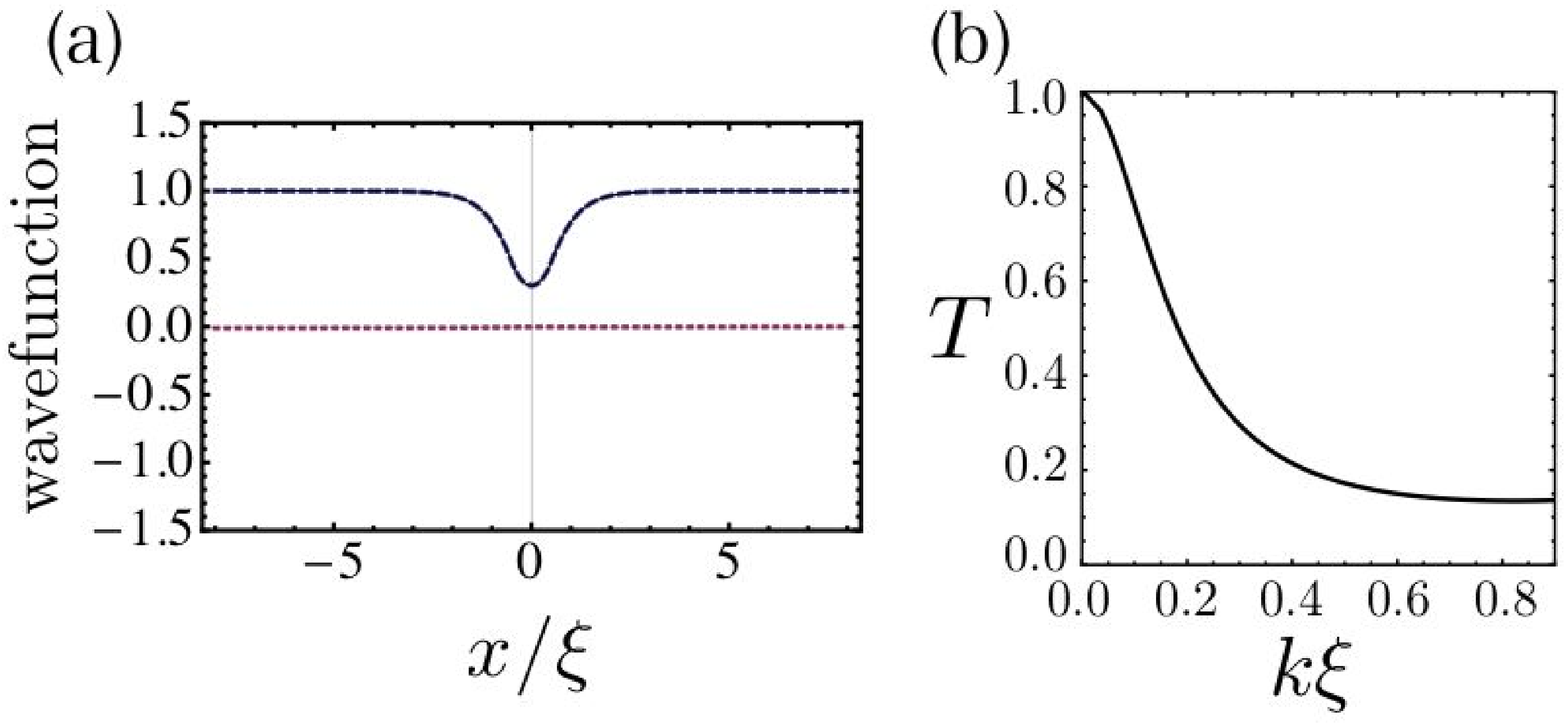}
\end{center}
\caption{(a) Wavefunctions as a function of the position $x$ in the ferromagnetic state at $B^{z} \neq 0$. 
The solid line represents the condensate wavefunction $\Phi_{+1}$. 
The dashed and dotted lines represent 
real and imaginary parts of the wavefunction of the spin wave mode in the long wavelength limit 
with energy $E = 10^{-7}c_{0}n + g\mu_{\rm B}B^{z}$, 
where the dashed line coincides with the solid line. 
We set $g\mu_{\rm B}B^{z} = 2 c_{0}n$.
(b) Transmission coefficient of the spin wave mode in the ferromagnetic state at $B^{z} \neq 0$ 
with $g\mu_{\rm B}B^{z} = 2 c_{0}n$. 
Although this excitation spectrum has the energy gap due to the Zeeman shift, this mode shows the total transmission in the long wavelength limit. } 
\label{fig5}
\end{figure}
One finds that all excitations but the quadrupolar spin mode in the ferromagnetic state 
show the total transmission through a potential barrier in the long wavelength limit. 
In striking contrast to these results, 
the total reflection occurs in the quadrupolar spin mode. 

There are several discussions about the total transmission of the Bogoliubov excitation within the spinless BEC~\cite{Kagan2003,Danshita2006,Kato2007,Ohashi2008}.  
The first interpretation was presented on the basis of a quasiresonant scattering 
near virtual level formed at zero energy~\cite{Kagan2003}. 
However, this interpretation was critically discussed comparing the wavelength of the excitation 
with the healing length~\cite{Danshita2006}. 
Other interpretations were submitted focusing on wavefunctions of excitations~\cite{Danshita2006,Kato2007,Ohashi2008}. 
In this section, we discuss a common property of excitations in BECs 
which show the total transmission in the long wavelength limit, on the basis of the idea discussed in Ref. 9.

Through a study of the total transmission of excitations in the spinless BEC, 
one of the authors and his collaborators found that the perfect transmission occurs 
when the wavefunction of the Bogoliubov excitation corresponds to that of the condensate 
in the low energy limit~\cite{Kato2007}. 
In the spinor BEC, 
we confirmed that this property also holds in excitations studied in the present paper. 
Wavefunctions of excitations in the spin-1 BEC which show the total transmission in the long wavelength limit 
correspond to that of the condensate. 

Figure 5 (a) shows the wavefunction of the spin wave in the ferromagnetic state, 
where the total transmission occurs in this mode. 
One can confirm that the wavefunction of the excitation with a long wavelength 
coincides with the condensate wavefunction. 
On the other hand, we confirmed that the wavefunction of the quadrupolar spin mode in the ferromagnetic state 
has a form different from that of the condensate in the long wavelength limit, 
where this mode shows the total reflection in this limit. 
The coincidence and the discrepancy of wavefunctions between the condensate and the excitations 
are crucial for the total transmission and the total reflection in the long wavelength limit.

We also note that the gapless spectrum and dispersion linear in $k$ are not always necessary to the total transmission. 
We have shown the total transmission both 
in the Bogoliubov excitation in the long wavelength limit (linear in $k$) (Fig.1 (a)) 
and in the excitation quadratic in $k$ (Fig.1 (b)). 
Although the spin wave mode in the ferromagnetic state has 
the energy gap due to the Zeeman shift at $B^{z} \neq 0$, 
this mode still shows the total transmission as shown in Fig. 5 (b). 

From those results, 
we shall make a conjecture, 
that {\it 
an excitation, whose wavefunction coincides with the condensate wavefunction in the long wavelength limit, 
transmits perfectly against a potential barrier in this limit, irrespective of properties of the excitation. 
Otherwise, reflection occurs}. 
We should remark that this is valid 
when the condensate has equal densities at $x = \pm \infty$. 
It is known that a partial transmission of the Bogoliubov excitation occurs, 
if we consider a case where 
densities of the condensate in asymptotic regimes are not equal. 
It is an interesting future problem to reveal how excitations transmit 
in such an asymmetric system in the spin-1 BEC.

\section{Conclusion}
We reported transmission properties of excitations in the spin-1 BEC. 
We found that all excitations but the quadrupolar spin mode in the ferromagnetic state 
experience the total transmission in the long wavelength limit. 
A correspondence between the wavefunction of an excitation and that of the condensate 
plays an important role in the total transmission.

\section*{Acknowledgement}  
This work is supported by the Ministry of Education, Science, Sports and Culture, 
Grant-in-Aid for Scientific Research on Priority Areas 20029007, and also supported by Japan 
Society of Promotion of Science, Grant-in-Aid for Scientific Research (C), 21540352. 
S. W. thanks support from 
the Fujyu-kai Foundation, and GCOE for Phys. Sci. Frontier, MEXT, Japan in the last fiscal year, 
and Grant-in-Aid for JSPS Fellows (217751) in the current fiscal year.



\begin{thebibliography}{99}
%

\bibitem{Stenger1998}
J. Stenger {\it et al.}, 
Nature {\bf 396} 345 (1998).

\bibitem{Higbie2005}
J. M. Higbie {\it et al.}, 
Phys. Rev. Lett. {\bf 95} 050401 (2005). 

\bibitem{Sadler2006}
L. E. Salder {\it et al.}, 
Nature (London) {\bf 443} 312 (2006). 

\bibitem{Kovrizhin2001}
D. L. Kovrizhin, Phys. Lett. A {\bf 287} 392 (2001). 

\bibitem{Kagan2003}
Yu. Kagan, D. L. Kovrizhin, and L. A. Maksimov, Phys. Rev. Lett. {\bf 90} 130402 (2003). 

\bibitem{Danshita2006}
I. Danshita, N. Yokoshi and S. Kurihara, 
New J. Phys. {\bf 8} 44 (2006).

\bibitem{Kato2007}
Y. Kato, H. Nishiwaki, and A. Fujita, 
J. Phys. Soc. Jpn. {\bf 77} 013602 (2007). 

\bibitem{Watabe2008}
S. Watabe, and Y. Kato, Phys. Rev. A {\bf 78} 063611 (2008). 

\bibitem{Ohashi2008}Y. Ohashi, and S. Tsuchiya, Phys. Rev. A {\bf 78} 043601 (2008). 

\bibitem{Takahashi2009}
D. Takahashi, and Y. Kato, J. Phys. Soc. Jpn. {\bf 78} 023001 (2009). 

\bibitem{Ohmi1998}
T. Ohmi, and K. Machida, J. Phys. Soc. Jpn. {\bf 69} 1822 (1998). 

\bibitem{Ho1998}
Tin-Lun Ho, Phys. Rev. Lett. {\bf 81} 742 (1998). 

\bibitem{Saito2006}
H. Saito, Y. Kawagchi, and M. Ueda, Phys. Rev. Lett. {\bf 96} 065302 (2006).  

\bibitem{2004Ieda}
J. Ieda, T. Miyakawa, and M. Wadati, 
Phys. Rev. Lett. {\bf 93} 194102 (2004). 

\end{thebibliography}
\end{document}